\documentstyle[11pt,IAU207_pasp,twoside,psfig]{article}
\def\edcomment#1{\iffalse\marginpar{\raggedright\sl#1\/}\else\relax\fi}
\reversemarginpar

\begin{document}
\title{Surface Photometry of Star Clusters\\
   in the Dwarf Elliptical Galaxies NGC 185 and NGC 205}
\author{Sang Chul Kim, Myung Gyoon Lee}
\affil{Astronomy Program, SEES, Seoul National University\\
   Seoul 151-742, Korea}
\author{Doug Geisler}
\affil{Departamento de F{\'\i}sica, Grupo de Astronom{\'\i}a, Universidad
   de Concepci\'on, Casilla 160-C, Concepci\'on, Chile}
\author{Ata Sarajedini}
\affil{Astronomy Department, University of Florida, Gainesville, FL 32611-2055, USA}
\author{Taft E. Armandroff}
\affil{National Optical Astronomy Observatory,
   P.O. Box 26732, Tucson, Arizona 85726, USA}
\author{Gary S. Da Costa}
\affil{Mount Stromlo and Siding Spring Observatories, The Australian
   National University, ACT 2611, Australia}

\begin{abstract}
We present the surface photometry of star clusters 
in the nearby dwarf elliptical galaxies NGC 185 and NGC 205,
obtained from deep HST WFPC2 F555W ($V$) and F814W ($I$) images.
We have obtained surface brightness and color profiles
of six star clusters in NGC 185, seven star clusters in NGC 205, and
one recently discovered non-stellar object in NGC 205.
The surface brightness profiles of ten star clusters are fitted well
by the King model, and
those of four star clusters are fitted well by the power-law.
Three out of ten star clusters fitted well with King model show 
signs of tidal tails.

\end{abstract}

It is essential to understand the structures of star clusters
in studying the dynamical evolutions of star clusters.
There have been several studies on the structures 
of star clusters in nearby galaxies 
(LMC, SMC, Fornax dwarf spheroidal, WLM, M31, and NGC 5128).
However, there is to date no study of the structures of star clusters
in dwarf elliptical galaxies.
 
\begin{figure}
\centerline{
\psfig{figure=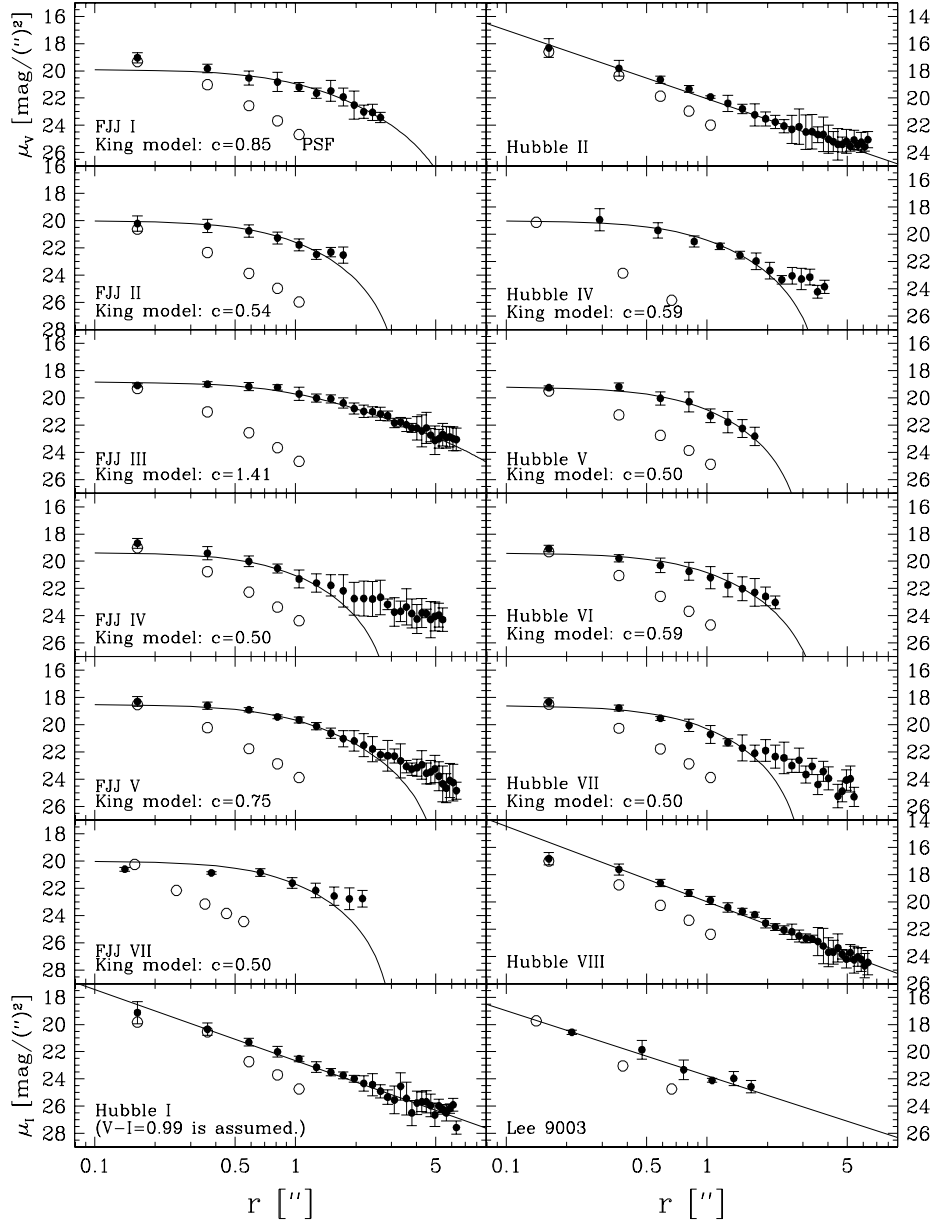,width=13cm,height=18cm}
}
\caption{F555W ($V$) surface brightness profiles
of the star clusters in NGC 185 and NGC 205 are shown with solid circles.
F814W ($I$) surface brightness profiles are shown for Hubble I.
Open circles represent point spread function(PSF) profiles.
Solid lines represent single-mass isotropic King models or 
power-law fit to the data.
Central concentration parameter $c$ is indicated for King model clusters.}
\end{figure}

We present, for the first time, the surface photometry results of 
star clusters in the nearby dwarf elliptical galaxies NGC 185 and NGC 205,
obtained from deep HST WFPC2 F555W and F814W images.
Surface photometry was obtained using concentric annular apertures.
Each concentric annular aperture was divided into eight sectors
to derive the mean surface brightness and error.

Surface brightness profiles of six star clusters in NGC 185
(FJJ I, FJJ II, FJJ III, FJJ IV, FJJ V and FJJ VII) and
eight star clusters in NGC 205 (Hubble I, Hubble II, Hubble IV, Hubble V,
Hubble VI, Hubble VII, Hubble VIII and Lee 9003) are shown
in Figure 1.  
Hubble I has only F814W band images due to spacecraft problems.
During his study of stellar populations in the central region of NGC 205,
Lee (1996) has noticed that Lee 9003 is a non-stellar object
in the CFHT CCD images.
Lee 9003 was covered in the WF4 chip of our Hubble V field,
and is found to be a star cluster candidate.

\begin{figure}
\centerline{
\psfig{figure=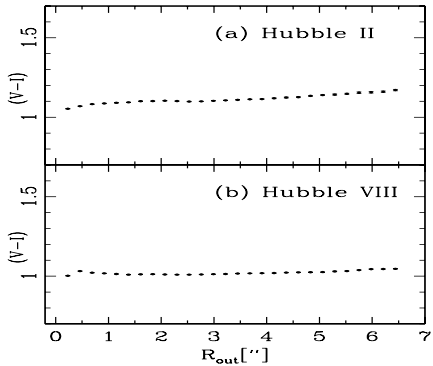,width=6cm,height=5cm}
}
\caption{Integrated color profiles of two selected star clusters
(a) Hubble II and (b) Hubble VIII in NGC 205.  }
\end{figure}

From Figure 1, it is found that seven star clusters
(FJJ I, FJJ II, FJJ III, FJJ VII, Hubble IV, Hubble V, and Hubble VI)
have surface brightness profiles well fitted by King (1966) model,
three star clusters (FJJ IV, FJJ V, and Hubble VII) have those with 
King model and tidal tails, and
four star clusters (Hubble I, Hubble II, Hubble VIII, and Lee 9003)
have those well fitted by power-law profiles.

Integrated color profiles are obtained for all the star clusters.
All the star clusters show little, if any, radial color gradient.
Figure 2 shows the color profiles of two selected star clusters (Hubble II and Hubble VIII).

Integrated photometry shows that the total magnitude ($M_V$) of the clusters 
range from $-5.6$ to $-8.3$ mag, adopting the distance modulus
 $(m-M)_0 = 23.96$ for NGC 185 and  $(m-M)_0 = 24.59$ for NGC 205.
Mean $M_V$ values for star clusters is derived to be 
$\overline{M_V} = -6.79 \pm 1.01$ mag for NGC 185, and
 $\overline{M_V} = -7.08 \pm 0.93$ mag for NGC 205
($-7.29 \pm 0.80$ mag without Lee 9003). 
This shows that the mean magnitudes of the star clusters
in NGC 185 and NGC 205 are somewhat fainter than those of our Galaxy.


\acknowledgments
This research is supported in part by the MOST/KISTEP 
International Collaboration Research Program (1-99-009).

\end{document}